\documentclass[11pt]{article}

\usepackage[margin=1in]{geometry}
\usepackage{booktabs}
\usepackage{hyperref}
\usepackage{enumitem}
\usepackage{xcolor}
\usepackage{graphicx}
\usepackage{CJKutf8}
\usepackage{amsmath}
\usepackage{amssymb}

\graphicspath{{figures_academic/}}

\hypersetup{
  colorlinks=true,
  linkcolor=black,
  urlcolor=blue,
  citecolor=black
}

\title{The Brain That Goes Quiet:\\
Serving a Large Model's Knowledge at 131 Tokens per Second\\
on an 8\,GB Laptop by Removing the Large Model from the Runtime Path}
\author{Myeong Jun Jo \\
ANIMA Research, Independent Researcher, South Korea \\
\texttt{ORCID: \href{https://orcid.org/0009-0006-9540-4666}{0009-0006-9540-4666}}}
\date{June 2026}

\begin{document}

\maketitle

\begin{abstract}
In earlier work I showed that a 35B-class Mixture-of-Experts model can be loaded and executed on a consumer laptop with 8\,GB of GPU memory. That result solved a placement problem and immediately exposed a different one: even correctly placed, the large model needed roughly four seconds to answer, because it was still being invoked at every query. This paper documents what happened when I stopped invoking it. During an offline phase, the large model reads source documents and writes verified answer entries into a structured knowledge store; at runtime, only a lightweight router, a deterministic renderer, and a 1B-class model are active. On the same 8\,GB laptop, end-to-end response time fell from approximately 4{,}465\,ms to 518\,ms, effective end-to-end throughput rose from 15.7 to 131 tokens per second, and the small model's streaming decode rate held at 226--237 tokens per second with a time-to-first-token of 29--62\,ms. The bottleneck turned out to be structural rather than model-specific: three different large models (Qwen, Gemma, and GLM class) all showed the same multi-second runtime cost, and all three produced usable knowledge stores offline. Two findings were less expected. First, on a 563-entry store built from seventeen real documents, simple keyword routing collapsed to 1.5\% top-1 accuracy while BM25-based routing reached 92.8\% (99.4\% top-3), and a confidence gate raised the effective top-1 to 98.0\% by escalating only 12.3\% of queries. Second, whether the small model preserves stored facts verbatim depends heavily on how the answer is packaged: across envelope formats carrying identical content, exact-match fidelity ranged from 9/9 to 0/9. A 16-case verification gate blocked all ten deliberately corrupted entries while admitting all six supported ones. The implementation internals are intentionally not disclosed; everything else, including the failure cases, is reported as observed.
\end{abstract}

\section{Introduction}

This paper continues a line of work that began with a deployment question rather than a research agenda. My professional background is in organizational operations and human resource management, and I currently work in an HR department. From that vantage point, the question that matters about large language models is rarely how to make them larger. It is how to make their capabilities usable inside ordinary organizations: places with procurement processes, closed networks, security policies, and laptops instead of accelerator clusters.

In a previous paper~\cite{paper16} I reported that a Qwen3.6-35B-A3B-class Mixture-of-Experts model could be placed and executed on a consumer laptop with an RTX 4060 Laptop GPU and 8\,GB of VRAM, using a rotary residency approach. I considered that a success, and in one sense it was: the model ran, and it ran stably. But anyone who actually sat in front of that laptop noticed the same thing I did. You typed a question, and then you waited. Roughly four seconds passed before the answer began to appear. The model was present, correctly placed, thermally stable --- and slow in a way that no placement scheme could fix, because the cost was not where the weights lived. The cost was the act of asking a 35B model anything at all at query time.

For a while I assumed the answer was incremental: trim the prompt, cache the prefix, shave the milliseconds. What changed my mind was not a serving paper but something I already had on my desk. In separate work, I had been developing a way for AI models to exchange information with each other in a machine-oriented format rather than in natural language --- originally intended for online communication between agents. One day it occurred to me that the same idea might apply here, in a direction I had not planned: if the large model could write its knowledge down in that format ahead of time, then at query time nobody would need to ask the large model anything. The moment I connected the two and the first answer came back fast and intact, my reaction was not surprise so much as confirmation. The judgment had been right.

The resulting system inverts the usual serving arrangement. The large model becomes an offline worker: it reads source material at night or during idle periods, produces answer candidates, and each candidate is checked against its source before being admitted into a structured knowledge store. At runtime, the large model is silent. A lightweight router maps the query to a stored entry, a deterministic renderer fixes the factual content, and a 1B-class model speaks it.

I call the system \emph{Horsehair} (\begin{CJK}{UTF8}{mj}연가시\end{CJK}), after the horsehair worm that steers the body of a mantis. The name is not decoration; it is an accurate description of the division of labor. The large model provides the body --- the knowledge, the judgment, the bulk. The small model is only the mouth. Getting the large model to actually stay quiet turned out to be one of the harder practical problems in this work, and I describe that experience honestly in Section~\ref{sec:failures}.

\paragraph{Contributions.} Concretely, this paper reports the following, all measured on a single 8\,GB consumer laptop:

\begin{enumerate}[leftmargin=*]
  \item Evidence that the multi-second runtime cost of large-model serving is structural and model-independent: three different large models (Qwen3.6-35B-A3B, Gemma4-26B-A4B, GLM-4.7-Flash class) all exhibit it in the runtime path, and all three produce usable knowledge stores offline (Section~\ref{sec:bottleneck}).
  \item An end-to-end comparison in which removing the large model from the runtime path reduces mean response time from 4{,}465\,ms to 518\,ms and raises effective end-to-end throughput from 15.7 to 131.4 tokens per second, with the small model's streaming decode rate at 226--237 tokens per second and time-to-first-token between 29 and 62\,ms (Section~\ref{sec:speed}). I deliberately report these three quantities --- end-to-end throughput, decode rate, and TTFT --- separately, because conflating them flatters the result.
  \item A routing study on a 563-entry store built from seventeen real documents (1{,}126 queries), where keyword routing collapses to 1.5\% top-1 while BM25-based routing reaches 92.8\% top-1 and 99.4\% top-3, and a margin-based confidence gate raises effective top-1 to 98.0\% by escalating 12.3\% of queries (Section~\ref{sec:routing}).
  \item An observation I did not anticipate: the small model's ability to relay stored facts verbatim depends on the packaging of the answer. Across envelope formats carrying identical content, exact-match fidelity ranged from 9/9 to 0/9, and a 232-token answer survived verbatim while a repetitive 236-token answer collapsed (Section~\ref{sec:fidelity}).
  \item A 16-case evaluation of the offline verification gate, which blocked all ten deliberately corrupted entries (fabrication, numerical error, logical reversal, contradiction, out-of-source claims) and admitted all six supported ones (Section~\ref{sec:gate}).
\end{enumerate}

The internal format of the structured store, the verification computation, and the stabilization machinery are deliberately not disclosed in this paper. Everything that is externally observable --- latencies, throughputs, accuracies, and failures --- is reported as measured, and the limitations section is written to be useful rather than ceremonial.

\section{Motivation: The Data Center Is Not the Only Way}

I want to be explicit about who this paper is for, because it shaped every decision in it. It is written for engineers inside ordinary organizations --- the people who are told that meaningful language-model capability requires either a cloud contract their security team will not approve, or a GPU cluster their budget will not cover. My experience in organizational operations is that this framing quietly ends most deployment conversations before they begin.

The premise of this work is that the framing is incomplete. A large model is expensive to run \emph{per query}, but most of what an organization asks of its internal knowledge systems is not novel reasoning per query. It is retrieval of judgments and facts that could have been produced once, carefully, in advance. If the expensive model does its thinking offline and the runtime path only routes, retrieves, and speaks, then the hardware requirement at the point of use collapses to something an ordinary laptop already has.

This is not an argument against data centers, any more than my previous paper was an argument against scaling. Frontier training and large-scale serving will remain where they are. The argument is narrower and, I think, more practical: for a meaningful class of organizational workloads, the large model's knowledge can be made to arrive in 30 milliseconds on 8\,GB of VRAM, and the only thing that has to be given up is the illusion that the large model itself must be awake when the question arrives.

\section{Related Work}

The components this work touches are individually well studied, and I want to position it carefully rather than claim novelty it does not have.

\paragraph{Caching and precomputation.} Semantic caching systems such as GPTCache~\cite{gptcache} store previous model responses and return them when a sufficiently similar query recurs. The present work shares the instinct that answers should be produced once and reused, but differs in three ways that matter in practice. First, entries here are not byproducts of past user queries; they are deliberately harvested offline from source documents, so coverage is a design decision rather than an accident of traffic. Second, every entry passes a source-grounded verification gate before admission, which a response cache by construction does not do. Third, the stored object is not free-form text but a structured representation whose packaging, as Section~\ref{sec:fidelity} shows, has a measurable effect on whether the serving model preserves it.

\paragraph{Retrieval-augmented generation.} RAG~\cite{rag} retrieves source chunks at query time and asks a generative model to compose an answer from them. The composition step is exactly what this architecture removes: here the answer was already composed and verified offline, and the runtime model is constrained to relay it. The trade is explicit --- RAG handles open-ended questions this system cannot, while this system offers determinism, auditability, and latency that generation-in-the-loop cannot.

\paragraph{Distillation.} Knowledge distillation~\cite{distill} moves a large model's behavior into a small model's weights. This work moves a large model's \emph{outputs} into an external store and leaves the small model generic. The store can be inspected, versioned, corrected entry by entry, and rebuilt overnight, none of which is true of distilled weights.

\paragraph{Fast inference.} Quantization, speculative decoding~\cite{speculative}, and MoE architectures~\cite{moe} accelerate the large model itself. My previous paper~\cite{paper16} sits in that tradition. The present paper takes the complementary path: rather than making the runtime invocation cheaper, it removes the invocation.

To my knowledge, the combination examined here --- offline harvesting by a large model into a verified, structured, deterministically renderable store, with the large model fully absent from the runtime path on consumer hardware --- has not been measured end to end in this form, and the envelope-fidelity result in Section~\ref{sec:fidelity} does not appear in the literature I am aware of. I would welcome corrections on both points.

\section{The Horsehair Architecture}

The architecture has two phases, and the easiest way to explain it is chronological, because that is how it was built.

\paragraph{Offline phase: the brain works at night.} The large model --- any of the three tested brains --- reads source material: in these experiments, a corpus of my own patent specifications and experiment logs, seventeen documents in total. For each knowledge item it produces an answer candidate together with keys and aliases for retrieval. Each candidate is then checked against the source that produced it. Candidates that survive are written into the structured store; candidates that fail are rejected. The exchange between the large model and the store does not use free natural language. It uses a machine-oriented structured format, developed in my earlier work on inter-model communication, that is simultaneously human-readable and deterministically parseable. That dual property is what makes the rest of the system possible: a human or a verifier can audit every entry, and the runtime can retrieve it with zero ambiguity. The internal details of the format and of the verification computation are not disclosed here.

\paragraph{Runtime phase: the brain goes quiet.} At query time, a router maps the incoming question to a stored key. A deterministic renderer produces the factual answer line from the stored entry --- this step involves no model and takes effectively zero time. A 1B-class model (MiniCPM5-1B, 4-bit) then streams the answer to the user. The small model is not asked to know anything or to compose anything; the factual content is locked by the renderer, and the small model's job is delivery. The large model is not loaded, not consulted, and not running.

\begin{figure}[h]
\centering
\includegraphics[width=0.6\linewidth]{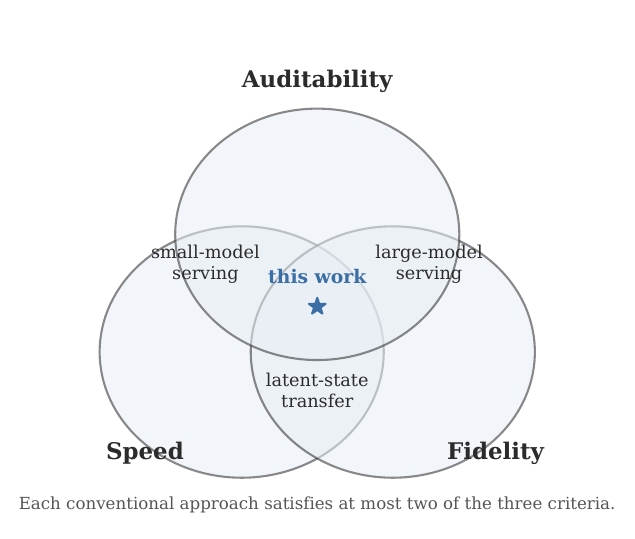}
\caption{The speed--fidelity--auditability trilemma in local serving. Small models are fast but drift; large models are faithful but slow; latent-state transfer is fast and faithful but opaque. The offline/runtime split targets all three.}
\label{fig:trilemma}
\end{figure}

\begin{figure}[h]
\centering
\includegraphics[width=0.85\linewidth]{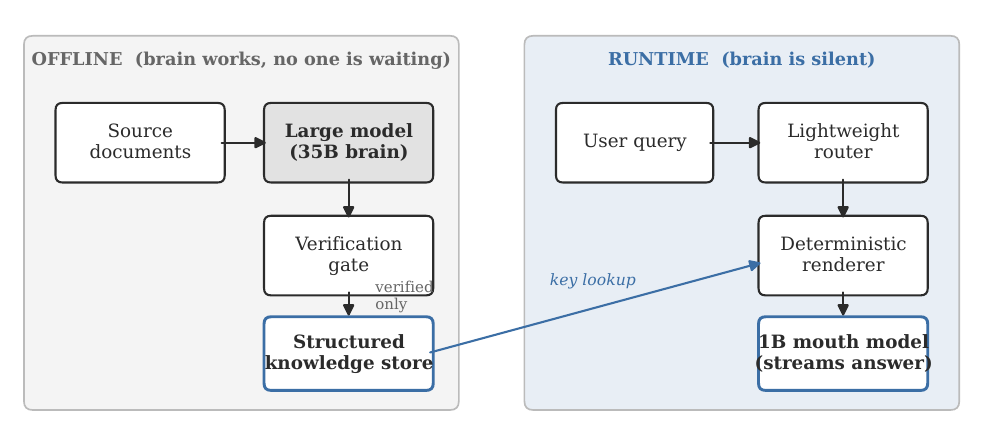}
\caption{Two-phase architecture. The large model is active only during offline harvesting; verified entries enter the structured store. At runtime, a router, a deterministic renderer, and a 1B-class mouth model serve the query. The large model is silent.}
\label{fig:architecture}
\end{figure}

Figure~\ref{fig:architecture} shows the two phases. Figure~\ref{fig:trilemma} states the design goal: existing local serving options give you at most two of speed, fidelity, and auditability, and the point of the offline/runtime split is to stop choosing.

\section{Experimental Setup}

All measurements were taken on a single consumer laptop: RTX 4060 Laptop GPU with 8\,GB VRAM, running ik\_llama.cpp. The offline brains were Qwen3.6-35B-A3B (placed via the rotary approach of~\cite{paper16}), Gemma4-26B-A4B, and GLM-4.7-Flash class models. The runtime mouth was MiniCPM5-1B at 4-bit quantization. Safety gates capped VRAM at 7{,}576\,MiB and GPU temperature at 78\,$^\circ$C throughout.

Latency experiments used five repeated runs per configuration ($N{=}5$), reported as means with min/max where relevant. The routing corpus contains 563 entries harvested from seventeen real documents, probed with 1{,}126 queries. The verification-gate evaluation uses 16 hand-labelled cases. Raw per-run CSV files, request/response logs, and harness scripts are retained as evidence for every number reported below.

\section{Results}

\subsection{The bottleneck is structural, and the brain is swappable}
\label{sec:bottleneck}

The first question was whether the four-second latency was a property of one model or of the arrangement. Table~\ref{tab:brains} answers it. Placed in the runtime path on this hardware, every large brain is slow in the same way: Qwen at 33.0 tokens per second of generation, Gemma at 20.8, GLM at 17.6, each behind roughly 4--7 seconds of wall-clock before a short answer completes. Meanwhile all three, used offline, produced valid structured store entries. The brain is interchangeable; the bottleneck is the act of invoking any of them at query time. This matters for deployment: an organization can swap the offline brain as better open models appear, without touching the runtime path.

\begin{table}[h]
\centering
\caption{Three large brains on the same task and hardware. All are bottlenecked in the runtime path; all produce valid store entries offline.}
\label{tab:brains}
\begin{tabular}{lrc}
\toprule
Model (offline brain) & Generation tok/s & Valid store entry \\
\midrule
Qwen3.6-35B-A3B & 33.0 & \checkmark \\
Gemma4-26B-A4B & 20.8 & \checkmark \\
GLM-4.7-Flash & 17.6 & \checkmark \\
\midrule
MiniCPM5-1B (runtime mouth) & 230 & --- \\
\bottomrule
\end{tabular}
\end{table}

\subsection{End-to-end speed}
\label{sec:speed}

Table~\ref{tab:latency} is the central comparison, and I want to be precise about definitions because this is where serving papers most often mislead. \emph{End-to-end throughput} divides the tokens of the final answer by total wall time, including routing, retrieval, HTTP overhead, and one prompt evaluation. \emph{Decode rate} is the small model's streaming generation speed once it begins. \emph{TTFT} is time to first token. These are different numbers and I report all three.

\begin{table}[h]
\centering
\caption{Runtime-path comparison, $N{=}5$ per configuration, same hardware and task. End-to-end throughput includes all overheads; decode rate is the streaming generation speed of the mouth model.}
\label{tab:latency}
\begin{tabular}{lrrr}
\toprule
Configuration & Mean wall time & End-to-end tok/s & Decode tok/s \\
\midrule
Qwen in runtime path & 4{,}464.6\,ms & 15.7 & 33.0 \\
Gemma in runtime path & 6{,}957\,ms & $\sim$9.8 & 20.8 \\
Store + 1B-router & 756\,ms & 90.1 & 226--237 \\
Store + keyword router & \textbf{518\,ms} & \textbf{131.4} & 226--237 \\
\bottomrule
\end{tabular}
\end{table}

With the large model removed, mean end-to-end wall time falls from 4{,}465\,ms to 518\,ms and end-to-end throughput rises from 15.7 to 131.4 tokens per second --- an 8.4$\times$ improvement measured honestly, with every overhead included. The mouth model's decode rate is 226--237 tokens per second, and TTFT once the server is warm is 29--62\,ms (53\,ms typical warm, 62\,ms on a cold cache; the one-time server load is 748\,ms). The residual gap between the 131 end-to-end figure and the 230 decode figure is HTTP overhead and a single prompt evaluation, and I report it rather than hiding it: closing that gap is engineering, not research.

I remember the run where the end-to-end number first crossed 100. That was the moment this stopped being an experiment about shaving milliseconds and became, to my mind, a statement about architecture: the four seconds had never been the model's fault. It was the cost of asking.

\begin{figure}[h]
\centering
\includegraphics[width=0.85\linewidth]{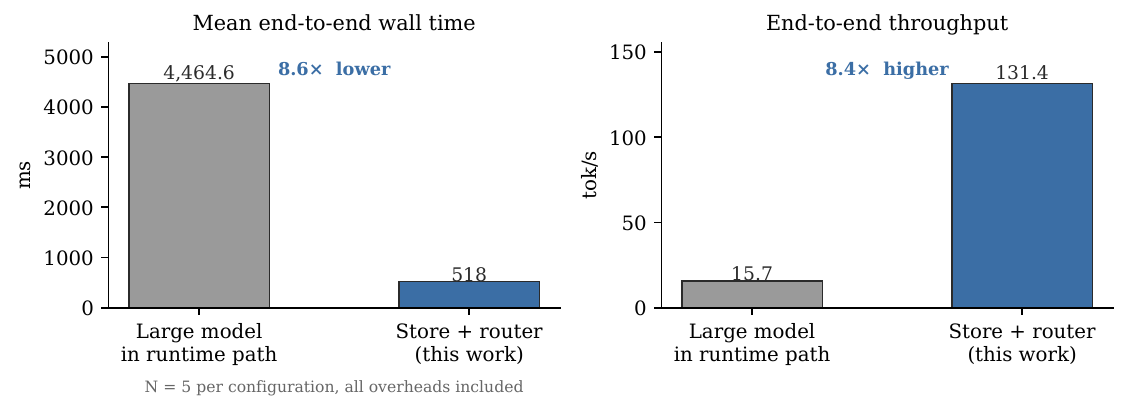}
\caption{Wall time and end-to-end throughput with the large model in and out of the runtime path. The improvement is 8.6$\times$ in wall time and 8.4$\times$ in end-to-end throughput, with all overheads included.}
\label{fig:latency-throughput}
\end{figure}

\subsection{Routing at realistic scale}
\label{sec:routing}

A three-key demonstration proves nothing about routing, so the store was rebuilt from seventeen real documents into 563 entries and probed with 1{,}126 queries. The result reshaped my view of the runtime path more than any other experiment.

\begin{table}[h]
\centering
\caption{Routing on the 563-entry store, 1{,}126 queries. Keyword routing, adequate on toy stores, collapses on real data.}
\label{tab:routing}
\begin{tabular}{lrrrr}
\toprule
Router & Top-1 & Top-3 & Wrong-key & Route p50 \\
\midrule
Keyword/regex & 0.015 & 0.046 & 0.493 & 2.45\,ms \\
BM25 & \textbf{0.928} & \textbf{0.994} & 0.072 & 2.40\,ms \\
Hybrid & 0.928 & 0.994 & 0.072 & 2.41\,ms \\
\bottomrule
\end{tabular}
\end{table}

Two things deserve emphasis. First, the keyword router that comfortably won the small-scale speed tests collapsed to 1.5\% top-1 accuracy on real documents --- a useful warning to anyone tempted to extrapolate from toy stores. Second, routing cost is negligible at this scale: a deliberately naive linear scan over 563 entries takes 2.4\,ms at the median, so the accuracy numbers above are not purchased with latency.

The error structure also turned out to be tractable. Correctly routed queries had a median score margin of about 0.30 over the runner-up; misrouted queries had a median margin near zero. A simple confidence gate exploits this: queries whose margin falls below a threshold ($\tau{=}0.05$) are escalated rather than answered. Only 12.3\% of queries trip the gate, and since 99.4\% of correct answers sit in the top-3, even a modest reranker on that escalated slice raises effective top-1 from 92.8\% to 98.0\%. The system, in other words, mostly knows when it is unsure --- which is precisely the property a no-hallucination serving path needs.

One caveat belongs here and not in fine print: the 1{,}126 queries were generated from distinctive vocabulary in the source documents, so 92.8\% is an optimistic bound. Free conversational phrasing routes worse, and I discuss that in the limitations.

\subsection{Fidelity depends on the envelope}
\label{sec:fidelity}

The result I least expected came from asking a mundane question: when the renderer hands the mouth model a locked answer, does the model actually repeat it? The answer is: it depends --- and what it depends on is not the content but the packaging.

I tested envelope formats that all carried byte-identical answer content and asked the mouth model to relay the answer across nine repetitions each. The spread was total. With the answer presented as a bare line following a structural marker, fidelity was 9/9 exact matches. With the same answer wrapped as a quoted string, 0/9. Wrapped as a JSON field, 0/9. The model does not treat packaging as neutral: quoting and JSON wrapping apparently invite it to interpret, translate, or summarize, while the marker-line format leaves it nothing to do but speak.

Length interacts with this but is not the driver. A 232-token non-repetitive answer survived verbatim (1.00 exact match at 204 tok/s); a 236-token answer containing repetitive text collapsed completely (0.00), with the model drifting into Chinese translation and paraphrase --- at one point rewriting ``VRAM slack'' as ``RAM slack,'' which in this domain is not a typo but a factual error. The practical rule adopted for the system is simple: answer lines of at most $\sim$230 tokens, no internal repetition, chunk anything longer.

I want to state the implication carefully. Within the conditions tested, a structured, marker-delimited representation was not a stylistic preference but a necessary condition for fact preservation through a small model. Natural-language-like packaging of the very same bytes destroyed fidelity. I had built the structured store for speed and auditability; the experiments suggest it was also, unknowingly, what kept the facts intact.

\begin{figure}[h]
\centering
\includegraphics[width=0.85\linewidth]{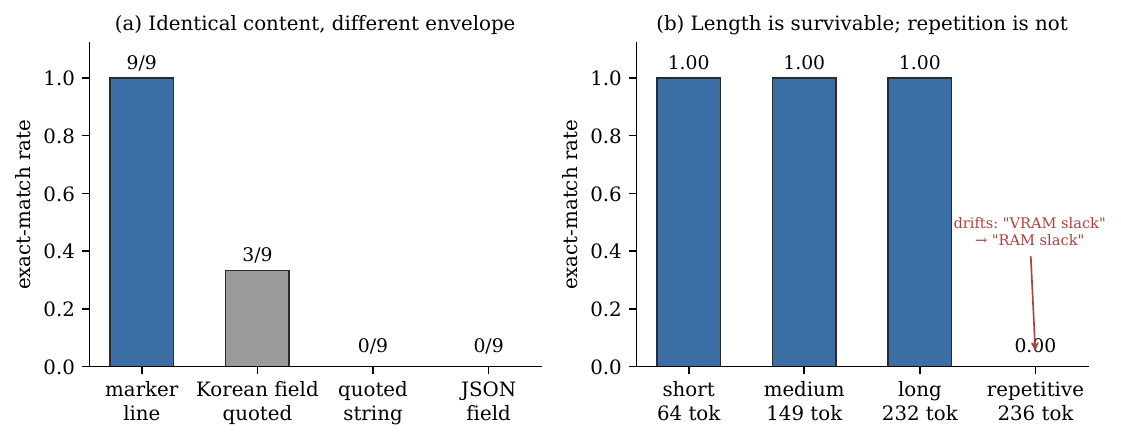}
\caption{Verbatim fidelity by answer packaging. Identical content yields 9/9 exact matches under a marker-line envelope and 0/9 under quoted-string and JSON envelopes. Repetitive text collapses fidelity regardless of length.}
\label{fig:ablation}
\end{figure}

\subsection{The verification gate}
\label{sec:gate}

Nothing enters the store unverified, so the gate itself needs evaluation. A 16-case labelled set was constructed: 6 answers genuinely supported by their sources, and 10 deliberately corrupted ones spanning fabrication, numerical error, logical reversal, internal contradiction, and out-of-source claims. The gate blocked all 10 corrupted cases and admitted all 6 supported ones --- precision, recall, and F1 of 1.0 on this set. Separately, harvesting from 8 real document chunks yielded 8/8 entries verified as supported.

Sixteen cases is a small set with relatively explicit errors, and I read this result as ``the gate is operational at its configured threshold,'' not as a generalization claim. Subtler paraphrase-level corruption would likely lower recall, and the verifier is itself a language model with the self-reference limits that implies.

\subsection{Context length is free}
\label{sec:context}

Because the runtime path never generates over a long context --- the heavy reading happened offline --- expanding the mouth model's context window should cost nothing at serve time. It does not. Sweeping the window from 4{,}096 to 131{,}072 tokens, decode rate stayed within 226--237 tokens per second (a 32$\times$ window expansion with no speed change), TTFT stayed between 29 and 48\,ms, and VRAM grew linearly at roughly 22\,MiB per 1k context, from 2.0\,GB at 4k to 5.0\,GB at 128k. At 262{,}144 the load itself exceeded the 7.6\,GB safety gate --- a VRAM ceiling, not a model ceiling. A 128k-context serving path fits comfortably inside an 8\,GB laptop, with headroom.

\begin{figure}[h]
\centering
\includegraphics[width=0.85\linewidth]{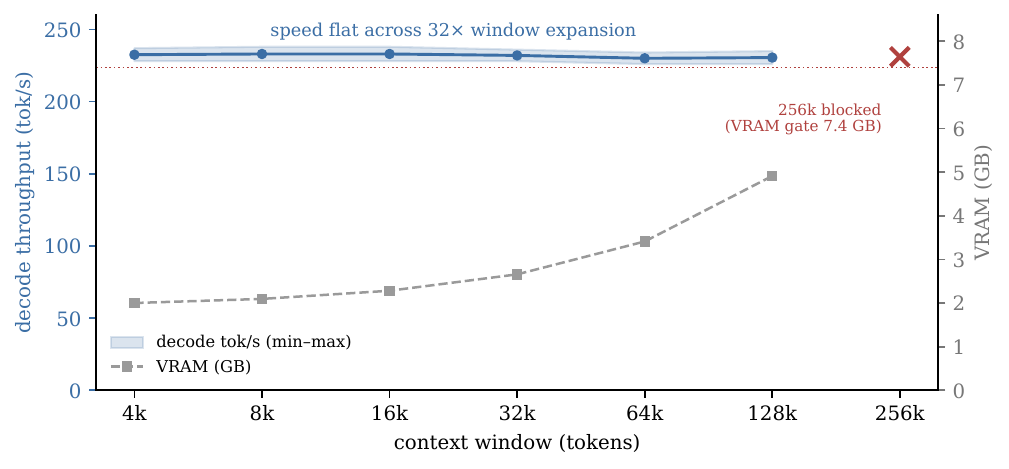}
\caption{Decode rate versus context window. Speed is flat from 4k to 128k because the large model never participates at runtime; the 256k point fails on VRAM, not on the model.}
\label{fig:context}
\end{figure}

\section{What Went Wrong Along the Way}
\label{sec:failures}

Papers usually present architectures as if they were designed in one sitting. This one was not, and three failures shaped it more than any design document.

The first was the latency itself --- the original motivation. The placed model from~\cite{paper16} worked, and waiting four seconds per answer made it feel broken anyway.

The second failure gave the system its name. In early configurations the large model, resident in its rotary placement, did not stay quiet: it kept participating in generation alongside the small model, burning tokens and dragging every response back toward large-model speed. The body kept trying to talk. Getting a strict division of labor --- the large model supplies, the small model speaks, never both --- took deliberate effort, and the final architecture enforces it absolutely by keeping the large model out of the runtime process entirely. The horsehair worm does not negotiate with the mantis.

The third was the hardware reminding me where I was working. Under 8\,GB, pushing the envelope did not produce graceful errors; it froze the screen. Those freezes are why every experiment here runs behind hard VRAM and temperature gates, and they reinforced the central design conviction: the safest large model on small hardware is one that is not running.

I include these not as color but because each one is a result. The second failure in particular --- that co-residency quietly degrades into co-generation --- is something anyone reproducing this class of system will meet, and I lost real time to it.

\section{Limitations}

\paragraph{Routing from free language.} The 92.8\% top-1 figure is measured on queries generated from document vocabulary. When users phrase questions colloquially, with synonyms or spelling variants, routing degrades, and a simple reranker did not consistently recover it. The confidence gate contains the damage --- low-margin queries are escalated rather than answered wrongly --- but a real fallback path for out-of-vocabulary phrasing (for example, having the offline brain generate paraphrase aliases per entry) is future work, not a solved problem.

\paragraph{Closed-world answers only.} The system answers what the store contains. It does not compose, infer, or handle novel questions; by design it refuses rather than improvises. Organizations whose query distribution is dominated by open-ended composition should use a different architecture, or this one only as a fast first tier.

\paragraph{Scale of evidence.} The verification gate is validated on 16 hand-labelled cases and 8 real harvested chunks. The full harvest-and-verify pipeline has not been run over thousands of documents, and its offline cost at that scale is unmeasured. The fidelity findings in Section~\ref{sec:fidelity} are from one mouth model; other small models may break differently.

\paragraph{Single machine, single user.} All numbers come from one laptop serving one user. Concurrency is unmeasured.

\paragraph{Disclosure boundary.} The structured format's internals, the verification computation, and the stabilization layer are not disclosed, which limits direct reproduction. The architectural claim, however, is testable without them: any deterministically parseable, source-verifiable store format should reproduce the shape of these results, and I would genuinely like to see someone try.

\section{Conclusion}

The four seconds were never the model's fault. They were the price of asking a large model questions at the moment they arrive, and this paper's contribution is the observation --- backed by measurements on an ordinary 8\,GB laptop --- that for a meaningful class of workloads, that price does not have to be paid at all. Move the asking offline, verify what comes back, lock it in a structured store, and the runtime collapses to a router, a renderer, and a small model that speaks at 230 tokens per second with a first token in tens of milliseconds. The large brain still does all the thinking. It just does it while no one is waiting.

Two findings travel beyond this system. Routing on real documents is harder than toy benchmarks suggest --- keyword matching collapsed outright --- but margin-based confidence makes the errors detectable rather than silent. And fact preservation through a small model is a property of the representation, not just the model: identical content survived or died by its packaging. Both points, I suspect, matter to anyone building local serving systems, whatever architecture they choose.

For the enterprise engineer this paper is addressed to, the message is short. The data center is not the only way. A laptop that holds a verified warehouse and a quiet brain can answer faster than the brain ever could aloud.

\vspace{2em}
\begin{center}
\begin{CJK}{UTF8}{mj}
\begin{tabular}{c}
踏雪野中去 \\
不須胡亂行 \\
今日我行跡 \\
遂作後人程 \\
\end{tabular}
\end{CJK}
\end{center}

\begin{center}
\small
\emph{Walking across a snowy field, \\
do not walk in disorder. \\
The tracks I leave today \\
become the path for those who follow.}
\end{center}

\begin{center}
\small attributed to Lee Yang-yeon; later cited by Kim Gu
\end{center}

\end{document}